\documentclass[12pt]{article}
\usepackage{amsmath}
\newlength{\extraspace}
\newlength{\extraspaces}
\newcommand{\be}{\begin{equation}
\addtolength{\abovedisplayskip}{\extraspaces}
\addtolength{\belowdisplayskip}{\extraspaces}
\addtolength{\abovedisplayshortskip}{\extraspace}
\addtolength{\belowdisplayshortskip}{\extraspace}}
\newcommand{\ee}{\end{equation}}
\newcommand{\ba}{\begin{eqnarray}
\addtolength{\abovedisplayskip}{\extraspaces}
\addtolength{\belowdisplayskip}{\extraspaces}
\addtolength{\abovedisplayshortskip}{\extraspace}
\addtolength{\belowdisplayshortskip}{\extraspace}}
\newcommand{\ea}{\end{eqnarray}}
\begin{document}
\thispagestyle{empty}
\begin{flushright}
SIT-LP-09/02 \\
February, 2009
\end{flushright}
\vspace{7mm}
%
%
\begin{center}
{\large{\bf Particle astrophysics in nonlinear supersymmetric \\[2mm]
general relativity}} 
\footnote{
\tt Talk given by K. Shima at the 4th EU RTN Workshop, 
{\it Constituents, Fundamental Forces and Symmetries of the Universe}, 
11-17 September 2008, Varna, Burgalia.} \\ [20mm]
%
{\sc Kazunari Shima}
\footnote{
\tt e-mail: shima@sit.ac.jp} \ 
and \ 
{\sc Motomu Tsuda}
\footnote{
\tt e-mail: tsuda@sit.ac.jp} 
\\[5mm]
{\it Laboratory of Physics, 
Saitama Institute of Technology \\
Fukaya, Saitama 369-0293, Japan} \\[20mm]
\begin{abstract}
An explanation of relations between the large scale structure of the universe 
and the tiny scale structure of the particle physics, 
e.g. the observed mysterious relation 
between the (dark) energy density and the dark matter of the universe 
and the neutrino mass and the SUSY breaking mass scale of the particle physics 
may be given by the nonlinear supersymmmetric general relativity (NLSUSY GR). 
NLSUSY GR shows that studying the physics before/of the Big Bang of the universe may be significant 
and may give new insight to unsolved problems of the low energy particle physics, 
cosmology and their relations. 
\\[5mm]
\end{abstract}
\end{center}

\newpage
\noindent
\section{Introduction}
Supersymmetry (SUSY) and Poincar\'e group are essential concepts for the unification of space-time and matter. 
We found that $SO(10)$ super-Poincar\'e (SP) group accomodates minimally the standard model (SM) 
with just three generations 
of quarks and leptons and the graviton in the {\it single} irreducible representation \cite{KS0}. 
We adopted the decomposition of $10$ supercharges as 
${\underline {10}}_{SO(10)}={\underline 5}_{SU(5)}+{\underline 5}^{*}_{SU(5)}$ 
according to $SO(10) \supset SU(5)$ 
and assigned the same quantum numbers as those of $\underline 5$ of $SU(5)$ GUT to ${\underline 5}_{SU(5)}$ 
satisfying $Q_{e}=I_{z}+{1 \over 2}(B-L)$. 
Regarding ${\underline 5}_{SU(5)}$ as a quintet of hypothetical spin-${1 \over 2}$ objects ({\it superon}) 
and all observed particles as composite eigenstates for space-time symmetry 
we have proposed the {\it superon-quintet model} (SQM) of matter, 
which gives potentially unified simple explanations about  the proton stability, various mixings of states, etc., 
though qualitative so far \cite{KS}. 
We discuss  in this article the field theoretical description of SQM including gravity, 
called {\it superon-graviton model} (SGM), 
and show some astroparticle consequences of SGM. 
SUSY SM can be regarded as an effective theory of SQM in asymptotic flat space-time 
describing the relativistic second order phase transition of massless SGM dictated 
by the global structure (symmetry) of space-time.

\section{Nonlinear supersymmetric general relativity \\
(NLSUSY GR)}
Toward the unified field theory of SQM including gravity, 
the supersymmetric coupling of spin-${1 \over 2}$ objects (superon) 
with spin-$2$ graviton is necessary.  
Nonlinear supersymmetric general relativity theory (NLSUSY GR) \cite{KSa}, 
which is based upon the general relativity (GR) principle 
and the nonlinear (NL) representation \cite{VA} of supersymmetry (SUSY) \cite{WZ,GL}, 
is the simple model and proposes a new paradigm called the SGM scenario \cite{KS,KSa,ST1,ST2} 
for the unified description of space-time and matter beyond (behind) the (SUSY) SM. 

The Volkov-Akulov (VA) model \cite{VA} is an almost {\it unique action} 
which represents {\it $N=1$ global SUSY nonlinearly} 
and describes the dynamics of massless  Nambu-Goldstone (NG) fermion of the spontaneous breakdown of space-time SUSY, 
i.e. the spontaneous SUSY breaking (SSB) mechanism is encoded {\it a priori} in the geometry of ultimate flat space-time. 
 (Alternatively, the geometry of ultimate flat (tangential) space-time is defined by the invariant volume action 
and induces self-contained phase transition to ordinary Minkowski space-time accompanying NG fermion, 
whose invariant action is given by the NLSUSY VA model \cite{VA}.) 

While, the NLSUSY model is recasted (related) to various linear (L) SUSY theories 
with SSB (abbreviated as {\it NL/L SUSY relation}), which has been shown by many authors 
in the various cases \cite{IK}-\cite{ST4}. 

In NLSUSY GR,  new (generalized) space-time, {\it SGM space-time} \cite{KS}, is introduced, 
where tangent space-time has the NLSUSY structure, 
i.e. flat space-time is specified not only by the $SO(3,1)$ Minkowski coodinates $x_a$ 
but also by $SL(2,C)$ Grassmann coordinates $\psi^i_\alpha$ ($i = 1, 2, \cdots, N$) for NLSUSY. 
The new Grassmann coordinates in  new (SGM) space-time mean 
coset parameters of ${super GL(4,R) \over GL(4,R)}$ 
which can be interpreted as the NG-fermions superon 
associated with the spontaneous breaking of super-$GL(4,R)$ down to $GL(4,R)$. 
The fundamental action in NLSUSY GR is given in the Einstein-Hilbert (EH) form in SGM space-time 
by extending the geometrical arguments of GR in Riemann space-time, 
which has a priori promising large symmetries 
isomorphic to $SO(10)$ ($SO(N)$) SP group \cite{ST1}. 

The SSB in NLSUSY GR due to the NLSUSY structure is interpreted 
as the phase transition of SGM space-time to Riemann space-time with massless superon (fermionic matter), 
i.e. {\it Big Decay} \cite{ST2,ST3} which subsequently ignites the Big Bang 
and the inflation of the present universe. 
In the SGM scenario all (observed) particles are assigned uniquely 
into a single irreducible representation of $SO(N)$ ($SO(10)$) SP group. 
And they are considered to be realized as (massless) eigenstates of $SO(N)$ SP 
composed of $N$ NG-fermions superon through the NL/L SUSY relation after Big Decay. 
As shown later the phase transition from the massless superon-graviton (SGM) phase to 
the composite eigenstates-graviton phase corresponds to the transition to the 
true vacuum (the minimum of the potential).
Since the cosmological term in NLSUSY GR gives the NLSUSY model \cite{VA} 
in {\it asymptotic} Riemann-flat (i.e. the ordinary vierbein $e^a{}_\mu \rightarrow \delta^a{}_\mu$) space-time, 
the SSB scale of NLSUSY, arbitrary so far, is now related to 
the Newton gravitational constant $G$ and the cosmological term of GR. 
As shown also later it induces (naturally) a fundamental mass scale 
depending on the cosmological constant 
and gives through the NL/L SUSY relation a simple explanation 
of the mysterious (observed) numerical relation 
between the (four dimensional) dark energy density of the universe and the neutrino mass \cite{ST2} 
in the vacuum of the $N = 2$ SUSY QED theory 
(in two-dimensional space-time ($d = 2$) for simplicity) \cite{STL,STLa}. 

In order to see the above particle physics consequences of  NLSUSY GR, i.e. 
{\it the relation between the large scale structure of space-time and the low energy particle physics}, 
let us begin with the fundamental EH-type action of NLSUSY GR in SGM space-time given by \cite{KSa} 
\begin{equation}
L_{\rm NLSUSYGR}(w) = {c^4 \over {16 \pi G}} \vert w \vert \{ \Omega(w) - \Lambda \}, 
\label{NLSUSYGR}
\end{equation}
where $G$ is the Newton gravitational constant, $\Lambda$ is a ({\it small}) cosmological constant, 
$\Omega(w)$ is the the unified scalar curvature 
in terms of the unified vierbein $w^a{}_\mu(x)$ (and the inverse $w_a{}^\mu$) defined by 
\begin{equation}
w^a{}_\mu = e^a{}_\mu + t^a{}_\mu(\psi), 
\ \ 
t^a{}_\mu(\psi) = {\kappa^2 \over 2i} 
(\bar\psi^i \gamma^a \partial_\mu \psi^i - \partial_\mu \bar\psi^i \gamma^a \psi^i), 
\label{unified-w}
\end{equation}
and $\vert w \vert = \det w^a{}_\mu$. 
In Eq.(\ref{unified-w}), $e^a{}_\mu$ is the ordinary vierbein of GR for the local $SO(3,1)$, 
$t^a{}_\mu(\psi)$ is the stress-energy-momentum tensor (i.e. the mimic vierbein) 
of the NG fermion $\psi^i(x)$ for the local $SL(2,C)$ 
and $\kappa$ is an arbitrary constant of NLSUSY with the dimemsion (mass)$^{-2}$. 
Note that $e^a{}_\mu$ and $t^a{}_\mu(\psi)$ contribute equally to the curvature of space-time, 
which may be regarded as the Mach's principle in ultimate space-time. 

The NLSUSY GR action (\ref{NLSUSYGR}) possesses promissing large symmetries 
isomorphic to $SO(N)$ ($SO(10)$) SP group \cite{ST1}; 
namely, $L_{\rm NLSUSYGR}(w)$ is invariant under \\[2mm]
%
[{\rm new \ NLSUSY}] $\otimes$ [{\rm local \ $GL(4,R)$}] 
$\otimes$ [{\rm local \ Lorentz}] $\otimes$ [{\rm local \ spinor \ translation}] \\
\hspace*{2.5cm} 
$\otimes$ [{\rm global}\ $SO(N)$] $\otimes$ [{\rm local}\ $U(1)^N$] $\otimes$ [{\rm Chiral}]. \\[2mm]
%
%
%
%
Note that the no-go theorem is overcome (circumvented) in a sense that 
the nontivial $N$-extended SUSY gravity theory with $N > 8$ has been constructed in the NLSUSY invariant way, 
i.e. by the degenerate vacuum (flat space-time).

\section{Particle astrophysics of NLSUSY GR}
New {\it empty} (SGM) space-time for {\it everything} described by the ({\it vacuum}) 
EH-type NLSUSY GR action (\ref{NLSUSYGR}) is unstable due to NLSUSY structure of tangent space-time 
and decays spontaneously to ordinary Riemann space-time with the NG fermions (superon matter) 
described by the ordinary EH action with the cosmological term, the NLSUSY action 
for the $N$ NG fermions and their gravitational interactions called SGM action.  
The SGM action is given by the following; 
\begin{equation}
L_{\rm SGM}(e,\psi) = {c^4 \over {16 \pi G}} e \vert w_{\rm VA} \vert \{ R(e) - \Lambda + T(e, \psi)\} 
\label{SGM}
\end{equation}
where $R(e)$ is the scalar curvature of ordinary EH action, 
$T(e,\psi)$ represents highly nonlinear gravitational interaction terms of $\psi^i$, 
and $\vert w_{\rm VA} \vert = \det w^a{}_b = \det (\delta^a_b + t^a{}_b)$ 
is the determinant in the NLSUSY model \cite{VA}. 
We can easily see that 
the cosmological term in $L_{\rm NLSUSYGR}(w)$ of Eq.(\ref{NLSUSYGR}) 
(i.e. the constant energy density of ultimate space-time) 
releted to the second term in the action (\ref{SGM}) reduces to the NLSUSY action \cite{VA}, 
$L_{\rm NLSUSY}(\psi) = -{1 \over {2 \kappa^2}} \vert w_{\rm VA} \vert$, 
i.e. the arbitrary constant $\kappa$ of NLSUSY is now fixed to 
\begin{equation}
\kappa^{-2} = {{c^4 \Lambda} \over {8 \pi G}} 
\label{kappa}
\end{equation}
in Riemann-flat $e_a{}^\mu(x) \rightarrow \delta_a^\mu$ space-time. 
Note that the NLSUSY GR action (\ref{NLSUSYGR}) and the SGM action (\ref{SGM}) possess 
different asymptotic flat space-time, i.e. 
SGM-flat $w_a{}^\mu \rightarrow \delta_a^\mu$ space-time 
and Riemann-flat $e_a{}^\mu \rightarrow \delta_a{}^\mu$ space-time, respectively. 
The scale of the SSB in NLSUSY GR (Big Decay of SGM space-time) produces a fundamental mass scale 
depending on the $\Lambda$ and $G$ through the relation (\ref{kappa}), 
which survives as the evidence of SGM scenario in the (low energy) particle physics in asymptotic flat space-time. 
Remember that the potential of (NLSUSY) $L_{\rm SGM}(e,\psi)$ (\ref{SGM}) (massless superon-graviton) is $\Lambda>0$.  

The nonlinear model sometimes can be converted (transformed) to the (equivalent) linear theory which is tractable 
and SUSY is also the case. 
It is interesting and important to investigate the low energy physics of NLSUSY GR 
through the NL/L SUSY relation. 
To see the (low energy) particle physics content in asymptotic Riemann-flat space-time 
we focus on $N=2$ SUSY in two dimensional space-time for simplicity, 
for in the SGM scenario $N=2$ case gives the minimal and realistic $N = 2$ LSUSY model \cite{STT}. 
By performing the systematic arguments for the NL/L SUSY relation we find the relation 
between the  $N=2$  NLSUSY model and a LSUSY QED theory \cite{ST3,ST4}; namely, 
\begin{equation}
L_{N=2{\rm SGM}}(e,\psi) \overset{e^a{}_\mu \rightarrow \delta^a_\mu}{\longrightarrow} 
L_{N=2{\rm NLSUSY}}(\psi) 
= L_{N=2{\rm SUSYQED}}({\bf V},{\bf \Phi}) + [{\rm tot.\ der.\ terms}]. 
\label{NLSUSY-SUSYQED}
\end{equation}
In the relation (\ref{NLSUSY-SUSYQED}), the $N = 2$ NLSUSY action $L_{N=2{\rm NLSUSY}}(\psi)$ 
for the two (Majorana) NG-fermions superon $\psi^i$ $(i = 1, 2)$ is written in $d = 2$ as follows; 
\begin{eqnarray}
&\!\!\! &\!\!\! 
L_{N=2{\rm NLSUSY}}(\psi) 
\nonumber \\
&\!\!\! &\!\!\! 
\hspace*{5mm} = -{1 \over {2 \kappa^2}} \vert w_{\rm VA} \vert 
= - {1 \over {2 \kappa^2}} 
\left\{ 1 + t^a{}_a + {1 \over 2!}(t^a{}_a t^b{}_b - t^a{}_b t^b{}_a) 
\right\} 
\nonumber \\
&\!\!\! &\!\!\! 
\hspace*{5mm} = - {1 \over {2 \kappa^2}} 
\bigg\{ 1 - i \kappa^2 \bar\psi^i \!\!\not\!\partial \psi^i 
- {1 \over 2} \kappa^4 
( \bar\psi^i \!\!\not\!\partial \psi^i \bar\psi^j \!\!\not\!\partial \psi^j 
- \bar\psi^i \gamma^a \partial_b \psi^i \bar\psi^j \gamma^b \partial_a \psi^j ) 
\bigg\}, 
\label{NLSUSYaction}
\end{eqnarray}
where $\kappa$ is a constant with the dimension $({\rm mass})^{-1}$, 
which satisfies the relation (\ref{kappa}) in the $d = 4$ case. 

On the other hand, in Eq.(\ref{NLSUSY-SUSYQED}), 
the $N = 2$ LSUSY QED action $L_{N=2{\rm SUSYQED}}({\bf V},{\bf \Phi})$ 
is constructed from a $N = 2$ {\it minimal} off-shell vector supermultiplet 
and a $N = 2$ off-shell scalar supermultiplet denoted ${\bf V}$ and ${\bf \Phi}$ respectively. 
Indeed, the most general $L_{N=2{\rm SUSYQED}}({\bf V},{\bf \Phi})$ in $d = 2$ 
with a Fayet-Iliopoulos (FI) $D$ term and Yukawa interactions, 
is given in the explicit component form as follows for the massless case; 
\begin{eqnarray}
L_{N=2{\rm SUSYQED}}({\bf V},{\bf \Phi}) 
&\!\!\! = &\!\!\!- {1 \over 4} (F_{ab})^2 
+ {i \over 2} \bar\lambda^i \!\!\not\!\partial \lambda^i 
+ {1 \over 2} (\partial_a A)^2 
+ {1 \over 2} (\partial_a \phi)^2 
+ {1 \over 2} D^2 
- {\xi \over \kappa} D 
\nonumber \\[.5mm]
& & 
+ {i \over 2} \bar\chi \!\!\not\!\partial \chi 
+ {1 \over 2} (\partial_a B^i)^2 
+ {i \over 2} \bar\nu \!\!\not\!\partial \nu 
+ {1 \over 2} (F^i)^2 
\nonumber \\[.5mm]
& & 
+ f ( A \bar\lambda^i \lambda^i + \epsilon^{ij} \phi \bar\lambda^i \gamma_5 \lambda^j 
- A^2 D + \phi^2 D + \epsilon^{ab} A \phi F_{ab} ) 
\nonumber \\[.5mm]
& & 
+ e \bigg\{ i v_a \bar\chi \gamma^a \nu 
- \epsilon^{ij} v^a B^i \partial_a B^j 
+ \bar\lambda^i \chi B^i 
+ \epsilon^{ij} \bar\lambda^i \nu B^j 
\nonumber \\[.5mm]
& & 
- {1 \over 2} D (B^i)^2 
+ {1 \over 2} A (\bar\chi \chi + \bar\nu \nu) 
- \phi \bar\chi \gamma_5 \nu \bigg\}
\nonumber \\[.5mm]
& & 
+ {1 \over 2} e^2 (v_a{}^2 - A^2 - \phi^2) (B^i)^2, 
\label{SQEDaction}
\end{eqnarray}
where $(v^a, \lambda^i, A, \phi, D)$ ($F_{ab} = \partial_a v_b - \partial_b v_a$) 
are the staffs of the minimal off-shell vector supermultiplet ${\bf V}$ 
representing $v^a$ for a $U(1)$ vector field, 
$\lambda^i$ for doublet (Majorana) fermions, 
$A$ for a scalar field in addition to $\phi$ for another scalar field 
and $D$ for an auxiliary scalar field, 
while ($\chi$, $B^i$, $\nu$, $F^i$) are the staffs of the (minimal) off-shell scalar supermultiplet ${\bf \Phi}$ 
representing $(\chi, \nu)$ for two (Majorana) fermions, 
$B^i$ for doublet scalar fields and $F^i$ for auxiliary scalar fields. 
Also $\xi$ in the FI $D$ term is an arbitrary dimensionless parameter turning to a magnitude of SUSY breaking mass, 
and $f$ and $e$ are Yukawa and gauge coupling constants with the dimension (mass)$^1$ (in $d = 2$), 
respectively. 
The $N = 2$ LSUSY QED action (\ref{SQEDaction}) can be rewritten in the familiar manifestly covariant form 
by using the superfield formulation (for further details see Ref.\cite{ST4}). 

In the relation (equivalence) of the two theories (\ref{NLSUSY-SUSYQED}), 
the component fields of $({\bf V},{\bf \Phi})$ in the $N = 2$ LSUSY QED action (\ref{SQEDaction}) are expanded systematically 
as composites of the NG fermions $\psi^i$, called {\it SUSY invariant relations} 
which terminate at ${\cal O}((\psi^i)^{4})$ (for the $d = 2$, $N = 2$ case), 
\begin{equation}
({\bf V}, {\bf \Phi}) \sim (\xi, \xi^i) \kappa^{n-1} (\psi^j)^n \vert w_{\rm VA} \vert + \cdots \ (n = 0,1,2), 
\label{SUSYinv}
\end{equation}
where $\xi^i$ is arbitrary demensionless ($SO(2)$) overall parameters 
in the SUSY invariant relations for ${\bf \Phi}$ 
and $(\psi^j)^2 = \bar\psi^j \psi^j, \epsilon^{jk} \bar\psi^j \gamma_5 \psi^k, 
\epsilon^{jk} \bar\psi^j \gamma^a \psi^k$, etc., 
which are very promissing and coinside with the SGM scenario. 
The explicit form \cite{ST3} of the SUSY invariant relations (\ref{SUSYinv}) are obtained {\it systematically} 
in the superfield formulation (for example, see Refs.\cite{IK,UZ,ST4}).  
And the familiar LSUSY transformations on the component fields of the LSUSY supermultiplets $({\bf V},{\bf \Phi})$ 
are reproduced in terms of the NLSUSY transformations on the constituents $\psi^i$. 
We find that {\it four-NG fermion self-interaction terms (i.e. the condensation of $\psi^i$)} 
appearing only in the auxiliary fields $F^i$ of the scalar supermultiplet ${\bf \Phi}$ 
is the origin of the familiar local $U(1)$ gauge symmetry of LSUSY theory \cite{ST3,ST4}. 
Is the condensation of NG-fermions superon the origin of the local $U(1)$ gauge interaction? 
The relation (\ref{NLSUSY-SUSYQED}) are shown explicitly (and systematically) 
by substituting Eq.(\ref{SUSYinv}) into the LSUSY QED action (\ref{SQEDaction}) \cite{ST3,ST4}. 

Now we briefly show the (physical) vacuum structure of $N = 2$ LSUSY QED action (\ref{SQEDaction}) 
related (equivalent) to the $N = 2$ NLSUSY action (\ref{NLSUSYaction}) \cite{STL,STLa}. 
The vacuum is determined by the minimum of the potential $V(A, \phi, B^i, D)$ in the action (\ref{SQEDaction}).    
%
%
The potential is  given by using the equation of motion for the auxiliary field $D$ as 
\begin{equation}
V(A, \phi, B^i) = {1 \over 2} f^2 \left\{ A^2 - \phi^2 + {e \over 2f} (B^i)^2 
+ {\xi \over {f \kappa}} \right\}^2 + {1 \over 2} e^2 (A^2 + \phi^2) (B^i)^2 \ge 0, 
\label{potential}
\end{equation}
The configurations of the fields corresponding to vacua $V(A, \phi, B^i)=0$ 
in $(A, \phi, B^i)$-space in the potential (\ref{potential}), 
which are dominated by $SO(1,3)$ or $SO(3,1)$ isometries, 
are classified according to the signatures of the parameters $e, f, \xi, \kappa$. 

By adopting the simple parametrization $(\rho, \theta, \varphi, \omega)$ for the vacuum configuration 
of $(A, \phi, B^i)$-space  and expanding the fields $(A, \phi, B^i)$ around the vacua 
we obtain the particle (mass) spectra of the linearized theory, $N = 2$ LSUSY QED. 
We have found that two different types of vacua $V(A, \phi, B^i)=0$ appear in the $SO(3,1)$ isometry \cite{STL,STLa} 
and that one of them is physical and describes $N=2$ LSUSY QED containing \\[2mm]
%
%
%
\hspace*{30mm} 
one charged Dirac fermion ($\psi_D{}^c \sim \chi + i \nu$), \\
\hspace*{30mm} 
one neutral (Dirac) fermion ($\lambda_D{}^0 \sim \lambda^1 - i \lambda^2$), \\
\hspace*{30mm} 
one massless vector (a photon) ($v_a$), \\
\hspace*{30mm} 
one charged scalar ($\phi^c \sim \theta + i \varphi$), \\
\hspace*{30mm} 
one neutral complex scalar ($\phi^0 \sim \rho {}(+ i \omega)$), \\[2mm]
with masses $m_{\phi^0}^2 = m_{\lambda_D{}^0 }^2 = 4 f^2 k^2 = -{{4 \xi f} \over \kappa}$,  
$m_{\psi_D{}^c} ^2 = m_{\phi^c}^2 = e^2 k^2 =-{{\xi e^2} \over {\kappa f}}$, 
$m_{v_{a}} = 0$,
%
which are the composites of NG-fermions superon 
and the vacuum 
breaks SUSY alone  spontaneously (The local $U(1)$ is not broken. For further detailes, see \cite{STL,STLa}). 

Remarkably these arguments show that the true vacuum of (asymptotic flat space-time of) $L_{N=2{\rm SGM}}(e,\psi)$ 
is achieved by the compositeness of fields (eigenstates) of the supermultiplet
of {\it global} $N = 2$ LSUSY QED. 
This phenomena may be regarded as the {\it relativistic} second order phase transition of massless superon-graviton system, 
which is dictated by the symmetry of space-time (analogous to the superconducting states achieved by the Cooper pair).

As for the cosmological significances of $N = 2$ SUSY QED in the SGM scenario, 
the (physical) vacuum for the above model explains (predicts) simply the observed mysterious (numerical) relation 
between {\it the (dark) energy density of the universe} $\rho_D$ ($\sim {{c^4 \Lambda} \over {8 \pi G}}$) 
and {\it the neutrino mass} $m_\nu$, 
\begin{center}
$\rho_D^{\rm obs} \sim (10^{-12} GeV)^4 \sim (m_\nu){}^4 
\sim {\Lambda \over G} \ (\sim {g_{\rm sv}}^2$), 
\end{center}
provided $- \xi f \sim O(1)$ and $\lambda_D{}^{0}$ is identified with  the neutrino, 
which gives a new insight into the origin of (small) mass \cite{ST2,STL,STLa} and 
produce the mass hielarchy by the factor ${e \over f}$($\sim O({m_{e} \over m_{\nu}})$ in case of $\psi_D{}^c$ as electron!). 

Furthermore, the neutral scalar field ${\phi^{0} (\sim  \rho) }$ with mass $\sim  O( m_\nu)$ 
of the radial mode in the vacuum configuration may be a candidate of {\it the dark matter}, 
for $N = 2$ LSUSY QED structure and the radial mode in the vacuum are preserved in the realistic large $N$ SUSY GUT model.  
(Note that $\omega$ in the model is a NG boson 
and disappears provided the corresponding local gauge symmetry is introduced as in the standard model.) 
%
%
The no-go theorem for $N > 8$ SUSY may be overcome in a sense that the linearized (equivalent) $N>8$ LSUSY theory would be 
{\it massive} theory with SSB.

Recently, by taking the more {\it general} auxiliary-field structure 
for the {\it general} off-shell vector supermultiplet \cite{ST5} and discussing the NL/L SUSY relation 
describing the phase transition to the true vacuum we have shown that 
{\it the magnitude of the bare (dimensionless) gauge coupling constant $e$} 
(i.e. the fine structure constant $\alpha = {e^2 \over 4\pi}$) is expressed (determined) in terms of {\it vacuum values 
of auxiliary-fields} \cite{ST5}: 
\begin{equation}
e_C = {\ln({\xi^i{}^2 \over {\xi^2 - 1}}) \over 4 \xi_C}, 
%
\label{f-xi}
\end{equation}
where $e_C$ is the bare gauge coupling constant, $\xi$, $\xi^i$ and $\xi_C$ are the vacuum-values (parameters) 
of auxiliary-fields of the {\it general} off-shell supermultiplets in $d = 2$.  
This mechanism is natural and very favourable for SGM scenario as a theory for everything. 
%

\section{Conclusions}
We have proposed a new paradigm for describing the unity of nature, 
where the ultimate shape of nature is new unstable space-time 
described by the NLSUSY GR action $L_{\rm NLSUSYGR}(w)$ in the form of 
the free EH action for empty space-time with the constant energy density. 
Big Decay of new space-time for $L_{\rm NLSUSYGR}(w)$ creates 
ordinary Riemann space-time with {\it massless} spin-${1 \over 2}$ superon 
described by the SGM action $L_{\rm SGM}(e,\psi)$ 
and ignites Big Bang of space-time and matter accompanying the dark energy (cosmological constant). 
Interestingly on Riemann-flat tangent space (in the local frame), 
the familiar renormalizable LSUSY theory emerges on the true vacuum of SGM $L_{\rm SGM}(e,\psi)$ 
as composite-eigenstates of superon. 
We have seen that the physics before/of the Big Bang may play crucial roles 
for understanding unsolved problems of the universe and the particle physics 
and for describing the rationale of being.  \par  
%
%
In fact, we have shown explicitly that $N=2$ LSUSY QED theory as the realistic $U(1)$ gauge theory  
emerges in the physical field configurations 
on the true vacuum of $N=2$ NLSUSY theory on Minkowski tangent space-time, which gives new insights into 
the origin of mass and the cosmological problems.  
The cosmological implications of the composite SGM scenario seem promissing but deserve further studies.   \par
Remarkably the physical particle states of $N=2$ SUSY as a whole 
look the similar structure to the lepton sector of ordinary SM with the local $U(1)$ 
and the implicit global $SU(2)$ \cite{STT} disregarding the R-parity. 
(Note that the scalar mode  $\omega$ is a NG boson and disappears provided the corresponding local gauge symmetry 
is introduced.) 
We anticipate that the physical cosequences obtained in $d=2$ hold in $d=4$ as well, 
for the both have the similar structures of on-shell helicity states of $N=2$ supermultiplet 
though scalar fields and off-shell (auxiliary field) structures are modified (extended). 
However, the similar investigations in $d = 4$ are urgent for the realistic model building based upon SUSY. \par  
The extension to large $N$, especially to $N = 5$ 
is important for {\it superon\ quintet\ hypothesis} of SGM scenario 
with ${N = \underline{10} = \underline{5}+\underline{5^{*}}}$ for equipping the $SU(5)$ GUT structure \cite{KS}  
and to  $N = 4$ may shed new light on the mahematical structures of 
the anomaly free non-trivial $d=4$ field theory. 
($N=10$ SGM predicts double-charge heavy lepton state $E^{2+}$ \cite{KS0}). 
Further investigations on the spontaneous symmetry breaking for $N \geq 2$ SUSY remains to be studied.  
It may be helpful to point out the formal similarity between superconductivity or superfluidity  and SGM scenario. \par
Linearizing SGM action ${L_{\rm SGM}(e,\psi)}$ on curved space-time, 
which elucidates the topological structure of space-time \cite{SK}, is a challenge. 
The corresponding NL/L SUSY relation will give the supergravity (SUGRA) \cite{FNF,DZ} 
analogue with the vacuum breaking SUSY spontaneously. \par  
Locally homomorphic non-compact groups $SO(1,3)$ and $SL(2,C)$ for  
space-time degrees of freedom are analogues of compact groups $SO(3)$ and $SU(2)$ for gauge degrees of freedom  
of 't Hooft-Polyakov monopole.
The physical and mathematical meanings of the black hole as a singularity of space-time and 
the role of the equivalence principle are to be studied in detail in NLSUSY GR and SGM scenario. \par
NLSUSY GR with extra space-time dimensions equipped with the Big Decay is also an interesting problem, 
which can give the framework for describing all observed particles as elementary {\it \`{a} la} Kaluza-Klein. 

Finally we just mention that NLSUSY GR and the subsequent SGM scenario 
for the spin-${3 \over 2}$ NG fermion \cite{ST1,Baak} is in the same scope. 
Our discussion shows that considering the physics before/of Big Bang may be significant for cosmology and 
the (low energy) particle physics as well. 
\vspace{1.5cm}

%
\noindent
One of the authors (K.S) would like to thank Professor D. Luest and Professor V. Dobrev for 
inviting him to the workshop and the warm hospitality during the stay at Varna. 
%
%
%

\newpage

%
\newcommand{\NP}[1]{{\it Nucl.\ Phys.\ }{\bf #1}}
\newcommand{\PL}[1]{{\it Phys.\ Lett.\ }{\bf #1}}
\newcommand{\CMP}[1]{{\it Commun.\ Math.\ Phys.\ }{\bf #1}}
\newcommand{\MPL}[1]{{\it Mod.\ Phys.\ Lett.\ }{\bf #1}}
\newcommand{\IJMP}[1]{{\it Int.\ J. Mod.\ Phys.\ }{\bf #1}}
\newcommand{\PR}[1]{{\it Phys.\ Rev.\ }{\bf #1}}
\newcommand{\PRL}[1]{{\it Phys.\ Rev.\ Lett.\ }{\bf #1}}
\newcommand{\PTP}[1]{{\it Prog.\ Theor.\ Phys.\ }{\bf #1}}
\newcommand{\PTPS}[1]{{\it Prog.\ Theor.\ Phys.\ Suppl.\ }{\bf #1}}
\newcommand{\AP}[1]{{\it Ann.\ Phys.\ }{\bf #1}}
\end{document}